\documentclass[aps,pre,showpacs,amsmath,amssymb,amsfonts,lengthcheck,twocolumn,longbibliography,superscriptaddress,floatfix]{revtex4-2}

\usepackage{lipsum}
\usepackage[T1]{fontenc}

\usepackage{changes} 
\usepackage{graphicx}
\usepackage{subfigure}
\usepackage{amsthm}
\usepackage{amsmath}
\usepackage{verbatim}
\usepackage{dcolumn}
\usepackage{bm}
\usepackage{color}
\usepackage{xcolor}
\usepackage{dsfont}
\usepackage{todonotes}
\usepackage{longtable}

\usepackage[colorlinks=true,citecolor=blue,linkcolor=blue,urlcolor=blue]{hyperref}

\allowdisplaybreaks

\newcommand{\bla}{bla\\bla\\bla\\bla\\bla}

\begin{document}

\title{Quantum Statistical Thermal Engine at the BCS--BEC crossover}

\author{Santiago Henríquez Lira}
\email{santiagohenriquezlira@gmail.com}
\affiliation{Instituto de Física, Pontificia Universidad Católica de Valparaíso, Casilla 4950, Valparaíso 2373223,
Chile}
\affiliation{Departamento de Física, Universidad Técnica Federico Santa María, Av. España 1680, Valparaíso 2390123, Chile}
\affiliation{Quantum Systems Unit, Okinawa Institute of Science and Technology Graduate University, Onna, Okinawa 904-0495, Japan}

\author{Felipe Isaule}
\affiliation{Departamento de Física, Universidad Técnica Federico Santa María, Av. España 1680, Valparaíso 2390123, Chile}

\author{Martín HvE Groves}
\affiliation{Instituto de Física, Pontificia Universidad Católica de Valparaíso, Casilla 4950, Valparaíso 2373223,
Chile}
\affiliation{Departamento de Física, Universidad Técnica Federico Santa María, Av. España 1680, Valparaíso 2390123, Chile}

\author{Francisco J. Peña}
\affiliation{Facultad de Ingeniería, Universidad San Sebastián, Lago Panguipulli 1390, Puerto Montt, Chile}

\author{Patricio Vargas}
\email{patricio.vargas@usm.cl}
\affiliation{Departamento de Física, Universidad Técnica Federico Santa María, Av. España 1680, Valparaíso 2390123, Chile}

\author{Thomás Fogarty}
\affiliation{Quantum Systems Unit, Okinawa Institute of Science and Technology Graduate University, Onna, Okinawa 904-0495, Japan}

\date{\today}

\begin{abstract}
We propose a quantum heat engine based on a two‑component Fermi gas with s‑wave contact interaction, operating across the BCS–BEC crossover. The work is extracted from the statistical properties of the gas, which are controlled by the interaction, rather than relying solely on compression and expansion stages. Using the functional renormalization group formalism, we obtain the equation of state in a non‑perturbative form along the crossover, encompassing the superfluid–normal phase transition. The cycle combines isentropic density strokes, isochoric thermalization, and isothermal interaction sweeps. This construction makes it possible to integrate features of both Otto and Carnot cycles, in which the system simultaneously saturates both efficiency limits without the net work vanishing. In the absence of density variations, the cycle reduces to a statistical Stirling‑like engine, in which the work generated arises exclusively from the
interaction, achieving efficiencies up to $38\%$. A pronounced asymmetry emerges through the crossover, giving rise to distinct operating regimes depending on the trajectory followed in the phase diagram. Consequently, the same architecture can be tuned to function as an engine, refrigerator, accelerator, or heater. These findings highlight pairing correlations as a versatile thermodynamic resource for quantum heat machines.
\end{abstract}

\maketitle

\section{Introduction}
\label{intro}
The performance of a quantum thermal machine is not necessarily fixed once its energy spectrum is chosen, as it can also depend on how the internal structure of the working medium is exploited during the cycle \cite{Dann2020,myers2022avs}. Several recent proposals have shown that this internal structure can itself be turned into a resource. Operating a many-body engine near a critical point, for instance, has been shown to reshape its performance beyond what compression and expansion alone can achieve \cite{campisi2016natcomm,fusco2016work,fogarty2021criticality,b2020universal}. Changing interactions in the working medium has also been shown to be a viable alternative for driving the engine cycle beyond the usual trap modulations \cite{chen2019npjqi,momo2023,nautiyal2024finite,watson2025quantum}. In ultracold-atom platforms, this idea has been realized experimentally by exchanging heat through individual, quantized atomic collisions with a thermal bath rather than through conventional thermal contact \cite{bouton2021natcomm}, and by driving a quasi-spin engine into a negative effective temperature to boost its output power \cite{nettersheim2022prxq}. Across these examples, the resource being exploited is not the spectrum itself, but the internal degrees of freedom of the working medium.

Quantum statistics is one such internal degree of freedom. Whether a working medium is described by Bose–Einstein or Fermi–Dirac statistics has been shown to reshape the thermodynamic response of an engine even when the underlying energy spectrum is held fixed \cite{zheng2015pre,Jaramillo_2016,Beau2016}. This effect has been made explicit by direct comparisons between bosonic and fermionic Otto cycles \cite{chen2018pre} and, more broadly, across quantum thermal machines built from either statistics \cite{sur2023entropy}. In particular, the permutation symmetry of bosons enables a collective enhancement with no fermionic counterpart \cite{watanabe2020prl}. This advantage has been shown to hold systematically across a wide range of operating conditions \cite{myers2020pre}.

Such bosonic advantage reaches its clearest form in Bose–Einstein condensates (BEC). In this direction, Ref.~\cite{Myers2022BEC} showed that a macroscopically occupied BEC ground state substantially enhances the work output of a quantum engine. Subsequently, within a thermodynamic–geometric framework describing processes in meso- and microscale systems driven by slow variations in external control~\cite{eglinton2023njp}, a geometric picture of BEC-induced power enhancement was provided for these types of cycles. Experimentally, an isentropic engine using a condensed bosonic Lithium gas confirmed significant gains in efficiency and power compared to a non-degenerate gas~\cite{simmons2023prr}. These works highlight how the interplay between quantum degeneracy and excitations in the thermal cloud can lead to large pressure differences, which can be exploited to enhance performance. 

Taken together, these results demonstrate that statistics can be manipulated to extract work. In this direction, ultracold atomic gases have emerged as a highly controllable platform for exploring the role of statistics. Among these, spin-1/2 Fermi gases offer a natural setting to change the statistics within a single cycle. Indeed, such gases feature the BCS-BEC crossover, where the physical nature of the relevant degrees of freedom evolves from weakly bound fermionic Cooper pairs on the BCS side to tightly bound bosonic molecules in the BEC limit~\cite{Zwerger2011BCSBEC,Randeria2014BCSBEC}. This crossover can be explored freely in ultracold atomic experiments by tuning the inter-atomic scattering length through Feshbach resonance techniques~\cite{chin2010feshbach}, as demonstrated in various laboratories over the years \cite{regal2004observation,bartenstein2004crossover, ku2012revealing}. 

These ideas were explored experimentally with ultracold atoms in Ref.~\cite{koch2023quantum}, which realized a cycle with statistical changes across the BCS-BEC crossover but no standard heat strokes. This experiment proved that such a cycle can operate as an engine, then named the Pauli engine, showing that statistics can be manipulated as a thermodynamic resource. Following these ideas, Ref.~\cite{menon2025leveraging} proposed a hybrid Otto-like engine that manipulates the statistics of a one-dimensional Lieb-Liniger gas to further increase performance. Similar ideas have also been recently explored with the working medium comprised of anyonic particles \cite{Dunlop2025,lal2026hybrid}, which showcases the growing number of systems that can be used to realize these new hybrid statistical cycles. 

Motivated by these developments, we propose a quantum thermal cycle that exploits the tunability of the BCS--BEC crossover as a thermodynamic resource. We introduce a Statistical Thermal Cycle (STC) in which two isothermal strokes at constant density are driven by changing the interaction parameter across the crossover. These strokes are referred to as fermionization and bosonization strokes, in the operational sense that the working medium is driven between regimes with predominantly fermionic and bosonic thermodynamic character. The cycle is completed by adiabatic and isochoric strokes, so that temperature, density, and interaction strength act as the relevant external control variables.

We examine the engine's performance across a wide range of parameters. These include interactions across the strongly interacting crossover regime, as well as temperature ranges such that the working medium crosses the superfluid-to-normal transition temperature during the cycle. To compute the thermodynamics across those regimes, we employ the functional renormalization group (FRG) method~\cite{wetterich1993exact,berges2002nonperturbative,dupuis2021ultracold}. This is a non-perturbative framework that has been proven successful in describing the BCS-BEC crossover ~\cite{scherer2011functional,boettcher2012ultracold,boettcher2014critical,diehl2010functional}. In particular, it enables us to extract the relevant thermodynamic quantities and identify the strokes without performing expensive numerical calculations.
   
The document is organized as follows. In Sec.~\ref{sec:model}, we present the microscopic model of the interacting Fermi gas, the FRG framework, and the thermodynamic reconstruction procedure used to obtain the relevant equation of state. In Sec.~\ref{sec:statistical_cycle}, we construct the statistical thermal cycle and derive the corresponding expressions for heat and work along each stroke. In Sec.~\ref{sec:results}, we analyze the engine's performance across the BCS–BEC crossover: we first isolate the contribution of the statistical strokes alone, then restore the compression and expansion stages, and identify the distinct thermodynamic operational regimes that the cycle can realize. Finally, in Sec.~\ref{sec:conclusions}, we summarize our main conclusions. Supporting thermodynamic data is provided in the Appendix.


\section{Model and thermodynamic framework}
\label{sec:model}

\subsection{Model and FRG ansatz}

This work considers a quantum engine where the working medium is a three-dimensional Fermi gas with two spin components. We focus on a balanced gas, with equal chemical potential $\mu$ and equal mass $m$ for both spins. By considering that the fermions interact via short-range $s$-wave attractive interactions, one can describe the system with a two-channel model~\cite{nagaosa1999quantum,Salasnich2016ZeroPoint}. This is described by the following microscopic action 
\begin{equation}
\begin{aligned}
    \mathcal{S}=&\int_0^\beta d\tau\int d^3x \Bigg[\sum_{\sigma=\uparrow,\downarrow}\psi^\dagger_\sigma\left(\partial_\tau-\frac{\nabla^2}{2m}-\mu\right)\psi_\sigma\\
    &+\phi^\dagger\left(\partial_\tau-\frac{\nabla^2}{4m}+\nu_\Lambda\right)\phi-g(\phi^\dagger\psi_\uparrow\psi_\downarrow+\text{H.c.})\Bigg],
\end{aligned}
    \label{eq:S}
\end{equation}
where $\tau$ is the imaginary time, with $\beta=1/T$ being the inverse temperature, and $x$ are coordinates over the three-dimensional space. The fields $\psi_\sigma$ and $\psi^\dagger_\sigma$ are Grassmann fields that represent fermions of spin $\sigma$, while $\phi$ and $\phi^\dagger$ are auxiliary bosonic dimer fields that represent pairs of fermions. The coupling $\nu_\Lambda$ is the detuning, which is tied to the scattering length $a$ for atom-atom collisions. Finally, $g$ is a Yukawa coupling that drives the atom-dimer interaction (see Ref.~\cite{Diehl2007FunctionalIntegral} for more details).  Note that we consider natural units, $\hbar=k_B=1$.

To compute the thermodynamics, which is necessary to examine the engine's performance, we construct a Legendre-transformed effective action $\Gamma$. At equilibrium, this is directly connected to the grand-canonical potential as $\Omega=\beta^{-1}\Gamma_{\text{eq}}$. We compute $\Gamma$ with the FRG approach. Within this method, one works in terms of the scale-dependent effective action $\Gamma_k$, where $k$ is a scale that suppresses lower-momentum fluctuations. 

In general, one works with a truncated ansatz for $\Gamma_k$. We use the ansatz used in Ref.~\cite{diehl2010functional}, which enables us to give a good description of the crossover above and below the superfluid phase transition~\cite{scherer2011functional}. The ansatz reads
\begin{equation}
\begin{aligned}
    \Gamma_k=&\int_0^\beta d\tau\int d^3x \Bigg\{\sum_{\sigma} \psi_\sigma^\dagger \left(\partial_\tau-\frac{\nabla^2}{2m}-\mu\right)\psi_\sigma\\
    &+\phi^\dagger\left(Z_{\phi,k}\partial_\tau-\frac{A_{\phi,k}}{4m}\nabla^2\right)\phi\\
    &+U_k(\rho)-h\left(\phi^\dagger\psi_\uparrow\psi_\downarrow+{\rm h.c.}\right)\Bigg\}\,,
\end{aligned}
\label{eq:effective_action_ansatz}
\end{equation}
where $\rho=\phi^\dagger\phi$, $Z_{\phi,k}$ and $A_{\phi,k}$ are renormalization factors, and $U_k$ is the effective potential. The latter is expanded as
\begin{equation}
\begin{aligned}
    U_k(\rho)=&m^2_{\phi,k}(\rho-\rho_0)+\frac{\lambda_k}{2}\left(\rho-\rho_{0,k}\right)^2\\
    &-(n_{k}+n_{1,k}(\rho-\rho_{0,k}))(\tilde{\mu}-\mu),
\end{aligned}
    \label{eq:potential_expansion}
\end{equation}
where $\rho_{0,k}$ is the minimum of $U_k$ and $\tilde{\mu}$ is a shift to the chemical potential. Here, a finite $\rho_{0,k}$ means the breaking of the U(1)-symmetry, signaling a superfluid gas.

The flow of $\Gamma_k$ as a function of $k$ is followed non-perturbatively by solving the FRG flow equation~\cite{wetterich1993exact}.  At high scales $k=\Lambda$, $\Gamma_k$ reduces to the known microscopic model~(\ref{eq:S}), and thus $\Gamma_\Lambda=\mathcal{S}$, which serves as initial condition. Therefore, the physical inputs of the calculations are the scattering length $a$, the chemical potential $\mu$, and the temperature $T$, which are tuned across the cycle. In turn, $\Gamma_k$ becomes the physical effective action $\Gamma$ at $k\to 0$, from which we can extract the thermodynamics. We refer to Ref.~\cite{diehl2010functional} for a detailed review of the calculation, including explicit expressions for the flow equations used in this work.

Importantly, $U_k$ is connected to the grand-canonical potential via $U_{k\to 0}=\Omega/V$, where $V$ is the volume. This enables us to compute the densities and entropies, as well as the internal energy. By identifying $n_{k}=-(\partial U_k/\partial \mu)_{\rho=\rho_0,\tilde{\mu}=\mu}$ as the flowing density, we can extract the physical density of the gas $n$ at $k\to 0$: $n=n_{k\to 0}$. Similarly, the entropy density $s$ is obtained at $k\to 0$ from $s_{k}=-(\partial U_k/\partial T)_{\rho=\rho_0,\tilde{\mu}=\mu}$. Here, we stress that the density $n$ is the control parameter used later in the compression and expansion strokes. In turn, the entropy is used to find isentropic strokes and calculate heat exchanges with the reservoirs.

Additionally, the internal energy is required to be able to compute the energy difference within strokes. To calculate this, we first calculate the pressure following the approach presented in Ref~\cite{isaule2020thermodynamics}. Firstly, we compute the pressure at zero temperature by integrating the density over the chemical potential as
\begin{equation}
    P(\mu,T=0)=\int_{\mu^{\text{(v)}}}^\mu n(\mu',T=0)\,d\mu',
    \label{eq:pressure_mu}
\end{equation}
where $\mu^{\text{(v)}}=\epsilon_b/2$. Here $\epsilon_b=-1/(ma^2)\Theta(a)$ is the two-body binding energy in the vacuum, where $\Theta$ is the Heaviside step function~\footnote{Note that we have used that the pressure vanishes in the vacuum, and thus $P(\mu^{\text{(v)}},T=0)=0$.}. The finite-temperature pressure is then obtained by integrating the entropy as
\begin{equation}
    P(\mu,T)=P(\mu,0)+\int_0^T s(\mu,T')\,dT'.
    \label{eq:pressure_T}
\end{equation}

After calculating the pressure, we can calculate the energy density $\epsilon=-P+\mu n+ T s$. In particular, to examine the performance of the cycle later in Sec.~\ref{sec:statistical_cycle}, we work in terms of the internal energy per particle
\begin{equation}
    e=\frac{\epsilon}{n}=-\frac{P}{n}+\mu+T\frac{s}{n}~~.
    \label{eq:internal_energy_per_particle}
\end{equation}
where $s/n=s^\star$ is the entropy per particle. These quantities enable us to fully compute the changes in the work and heat per particle $w$ and $q$, respectively, which dictate the engine's performance.

\subsection{Crossover physics}

One essential aspect of our engine is manipulating the interaction between spins. As mentioned, this interatomic interaction is characterized by the scattering length $a$, which can be controlled by Feshbach resonance techniques \cite{chin2010feshbach}. This scattering length is small and negative in the limit of weak attraction, where the fermions can form loosely bound Cooper pairs, and thus the gas becomes a BCS superfluid with fermionic statistics. In contrast, $a$ is small and positive in the limit of strong attraction, where the fermions form tightly bound dimers, forming a BEC with bosonic statistics. Both limits are continuously connected through the BCS-BEC crossover, including the unitary limit where $a\to\infty$. Therefore, by changing the scattering length, we can control the system's statistics at will. 

The crossover is characterized by the dimensionless interaction parameter
\begin{equation}
    c_F=\frac{1}{k_Fa}~~.
    \label{eq:cF}
\end{equation}
Here, $k_F$ is the Fermi momentum, which is defined as 
\begin{equation}
    k_F=(3\pi^2 n)^{1/3},
\end{equation}
where $n$ is the density. From $k_F$, we can also define the Fermi energy as $E_F=k_F^2/(2m)$.

The BCS fermionic limit is recovered for $c_F\ll -1$, the BEC bosonic limit for $c_F\gg 1$, while the crossover region lies around $c_F\sim 0$ \cite{Randeria2014BCSBEC}. Throughout this work, we perform cycles across the crossover with strokes that change the value of $c_F$ from positive to negative values, and vice versa.

\begin{table}[!b]
\caption{Summary of the main symbols used in the cycle analysis.}
\label{tab:notation}
\begin{ruledtabular}
\begin{tabular}{ll}
\small Symbol & \small Meaning \\
\hline\\[-1.6mm]
$c_F=1/(k_F a)$ & interaction parameter (dimensionless) \\
$c_F^{\ell},\,c_F^{h}$ & initial / target branches \\
$\theta=T/E_F$ & reduced temperature \\
$T_L,\,T_H$ & cold / hot reservoir temps \\
$\theta_L,\,\theta_H$ & reduced cold / hot temps \\
$n_1,\,n_2$ & densities (low / high) \\
$r_t=T_L/T_H$ & thermal ratio of the bath\\
$r_c=n_2/n_1$ & compression ratio \\
$e=\epsilon/n$ & energy per particle \\
$s^{\star}=s/n$ & entropy per particle \\
$w,\,q$ & work / heat per particle \\
$w^{s},\,q^{s}$ & statistical work / heat per particle \\
$w^{\rm int}$ & internal (non-mech.) work \\
$T_c^{\mathrm{BCS}},T_c^{\mathrm{BEC}}$ & critical temps (on each side) \\
$\textbf{A}$--$\textbf{F}$ & cycle states \\
\end{tabular}
\end{ruledtabular}
\end{table}


\section{Thermodynamic cycle}
\label{sec:statistical_cycle}

In this section, we introduce the Statistical Thermal Cycle (\textbf{STC}) for the working medium defined in Sec.~\ref{sec:model}. The cycle consists of six stages, or strokes, where the control variables are the scattering length $a$, the density $n$, and the temperature $T$. Here we stress that $n$ is found by fine-tuning the chemical potential $\mu$.

\begin{figure*}[t]
\centering
\includegraphics[width=1\textwidth]{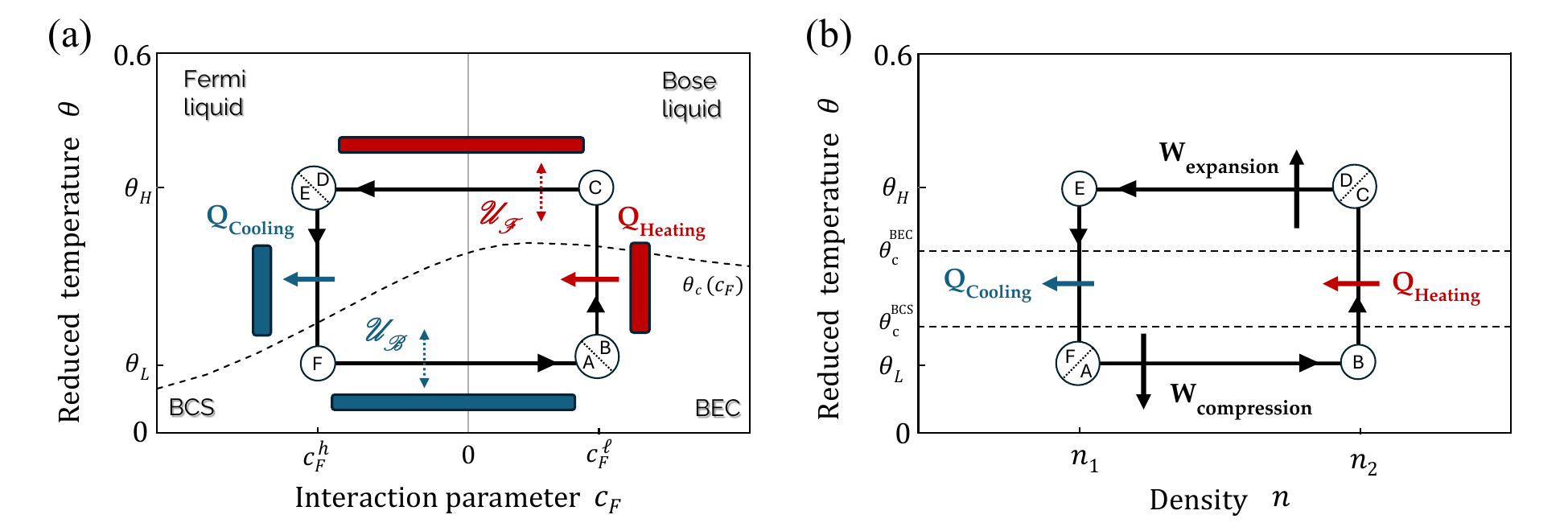}
\caption{ Illustration of the cycle at the BCS-BEC crossover, with $c_F=(k_Fa)^{-1}$ and $\theta=T/{E_F}$. The cycle consists of: adiabatic compression (\textbf{A--B}), isochoric heating (\textbf{B--C}), a hot statistical isotherm (\textbf{C--D}), adiabatic expansion (\textbf{D--E}), isochoric cooling (\textbf{E--F}), and a cold statistical isotherm (\textbf{F--A}). Strokes which include coupling to the hot and cold reservoirs, are indicated by the red and blue colors, respectively.
(a) Trajectory in the $(\theta, c_F)$ plane, showing the initial $(c_F^\ell)$ and target $(c_F^h)$ branches relative to the superfluid transition $\theta_c = T_c/E_F$ (dashed curve); $\mathcal{U}$ indicates the statistical stroke, where both heat and work are exchanged. (b) Trajectory in the $(\theta, n)$ plane, showing the density-changing strokes \textbf{A--B} and \textbf{D--E}, together with the critical temperatures $\theta_c^{BEC}$ and $\theta_c^{BCS}$.}
\label{fig:diagrama_stc}
\end{figure*}

In Table~\ref{tab:notation} we list the main symbols used to analyze each cycle, while in Table~\ref{tab:control_variables} the control parameters of each stroke are summarized, as well as whether heat and/or work is exchanged. Similar to the cycle reported in Ref.~\cite{menon2025leveraging}, the cycle includes two isentropic, two isochoric, and two statistical isothermal strokes, and is illustrated in Fig.~\ref{fig:diagrama_stc}. Since the working medium is controlled by the three variables mentioned above, the complete cycle describes a trajectory in the control space $(\theta, c_F, n)$. Fig.~\ref{fig:diagrama_stc}(a) shows the $(\theta, c_F)$ plane, illustrating the transition from bosonic statistics (right side of the figure) to fermionic statistics (left side of the figure) along the cycle, and its coupling to the hot and cold reservoirs which can realize the normal–superfluid transition, $\theta_c$ (dashed line). In turn, Fig.~\ref{fig:diagrama_stc}(b) shows the $(\theta, n)$ plane, highlighting the compression and expansion stages, without normal–superfluid phase changes along the stroke.

Throughout this work, each state $j$ is denoted by
\begin{equation}
    j=(\theta_j,c_{F,j})_{n_j}\,,~~~
    j=\textbf{A},\textbf{B},\textbf{C},\textbf{D},\textbf{E},\textbf{F}\,,
    \label{eq:state_notation_cycle}
\end{equation}
 where the subscript $n_j$ identifies the density of the gas in a given state and $\theta_j$ the reduced temperature, in which the associated temperature is
\begin{equation}
    T_j=E_F(n_j)~\theta_j\,.
    \label{eq:physical_temperature_state}
\end{equation}
 
The cycle is driven between the densities, $n_1$ and $n_2$, and operates between two temperatures, $T_L$ and $T_H$, representing two thermal baths at low and high temperatures, respectively. Consequently, according to Eq.~\eqref{eq:physical_temperature_state}, the reduced reservoir temperatures are defined as
\begin{equation}
   \theta_{L}=\frac{T_L}{E_{F,1}}\,,~~~
    \theta_{H}=\frac{T_H}{E_{F,2}}\,.
    \label{eq:reduced_bath_temperatures}
\end{equation}

As shown in  Fig.~\ref{fig:diagrama_stc} and deduced from the information in Table~\ref{tab:control_variables}, it follows that the cycle operates between two interaction parameters,
\begin{equation}
c_F^{\ell} = c_{F,\textbf{A}} = c_{F,\textbf{B}} = c_{F,\textbf{C}}\,,
\label{eq:cF_low_branch}
\end{equation}
and 
\begin{equation}
c_F^{h} = c_{F,\textbf{D}} = c_{F,\textbf{E}} = c_{F,\textbf{F}}\,.
\label{eq:cF_high_branch}
\end{equation}
The states in Eq.~\eqref{eq:cF_low_branch} are referred to as the initial branch represented by $c_F^{\ell}$, whereas the states in Eq.~\eqref{eq:cF_high_branch} correspond to the target branch represented by $c_F^{h}$.

\begin{table}[!t]
\caption{Control variables, fixed parameters, and exchanged quantities with the environment during each stroke.}
\label{tab:control_variables}
\begin{ruledtabular}
\begin{tabular}{lccc}
Stages & Control & Fixed & Exchange \\
\hline\\[-1.6mm]
Compression/Expansion & \(n,a\) & \(s^\star, c_F\) & \(w\) \\
Heating/Cooling & \(T\) & \(n, c_F\) & \(q~\) \\
Statistical Isotherm H/C & \(a\) & \(T, n\) & \(q^s,\ w^s\)
\end{tabular}
\end{ruledtabular}
\end{table}

\subsection{Statistical Thermal Cycle}
\label{subsec:STC}

The cycle starts in the BEC regime with $c_{F,\textbf{A}}=c_F^\ell$, $\theta_\textbf{A}=\theta_{L}$ and $n_\textbf{A}=n_1$. Therefore, the state according to Eq.~(\ref{eq:state_notation_cycle}) is defined as
\begin{equation}
    \textbf{A}=\left(\theta_L~,~c_F^{\ell}\right)_{n_1}\,.
    \label{eq:state_A_cycle}
\end{equation}
The six strokes are defined as follows.

\subsubsection*{\texorpdfstring{\textbf{1. Adiabatic compression \(\textbf{\textup{A}}\to \textbf{\textup{B}}\)}}{\textbf{1. Adiabatic compression A->B}}}

The isolated gas is compressed from \(n_1\) to \(n_2\) at fixed \(c_F^{\ell}=c_{F,\textbf{B}}\). The final state is
\begin{equation}
\textbf{B}=\left(\theta_\textbf{B}~,~c_F^{\ell}\right)_{n_{2}}\,.
    \label{eq:state_B_cycle}
\end{equation}
During this stroke, the Fermi energy changes from $E_{F,\textbf{A}}$ to $ E_{F,\textbf{B}}$ proportionally to the change in temperature from $T_\textbf{A}$ to $T_\textbf{B}$. Thus, $\theta_\textbf{A} = \theta_\textbf{B}$. The temperatures satisfy the following condition
\begin{equation}
  T_{L}(n_{1}) <T_\textbf{B}(n_{2}) < T_c(n_{2})\,, 
\end{equation}
where $T_c$ is the critical condensation temperature in the high density branch $n_2$ and \(T_\textbf{B}\) is fixed by the isentropic condition
\begin{equation}
    s^\star_\textbf{B}(T_\textbf{B},c_F^{\ell})
    =
    s^\star_\textbf{A}(T_L,c_F^{\ell})\,.
    \label{eq:isentropic_AB}
\end{equation}
Because \(q_\textbf{AB}=0\), only work is done during this stroke, which is calculated from
\begin{equation}
\begin{aligned}
    w_\textbf{AB}
    &=
    e_\textbf{B}(T_\textbf{B},c_F^\ell) - e_\textbf{A}(T_L,c_F^\ell)\,.
\end{aligned}
\label{eq:wAB}
\end{equation}

\subsubsection*{\texorpdfstring{\textbf{2. Isochoric heating} \(\textbf{\textup{B}}\to \textbf{\textup{C}}\)}{\textbf{2. Isochoric heating B->C}}}

From state $\textbf{B}$, with the density fixed at \(n_2\) and the interaction remaining unchanged \(c_F^{\ell}=c_{F,\textbf{C}}\), the gas is coupled to the hot reservoir at temperature $T_{H}$ until equilibrium is reached. The final state is
\begin{equation}
    \textbf{C}=\left(\theta_H~,~c_F^{\ell}\right)_{n_2}
    \label{eq:state_C_cycle}\,.
\end{equation}
During this process, the temperature increases continuously from $T_\textbf{B}$ to $T_{H}$. However, due to correlations, the absorbed energy contributes not only to the increase in the total internal energy $e_\textbf{BC}$, but also to pair dissociation which leads to a simultaneous decrease in the correlation energy. As a result, the average entropic temperature is lower than that of the reservoir. Therefore, we must define the exchanged heat unambiguously as the entropic integral of the bath, and separate it from the total change in the reconstructed internal energy as
\begin{equation}
    \Delta e_\textbf{BC}=q_\textbf{BC}+w_\textbf{BC}^{\rm int}\,.
\end{equation}
Thus, the effective absorbed heat $q_\textbf{BC}$ is evaluated as
\begin{equation}
  q_\textbf{BC}=\frac{1}{n_2} \int_{s_\textbf{B}}^{s_\textbf{C}} T\,ds=\frac{1}{n_2} \int_{T_\textbf{B}}^{T_\textbf{C}} T\,\left( \frac{\partial s}{\partial T} \right)_{n_2, c_F^\ell}dT\,.
   \label{eq:qbc}
\end{equation}
The mean entropic temperature of this stroke is then given by
\begin{equation}
    \overline{T}_\textbf{BC} =\frac{q_\textbf{BC}} {(s_\textbf{C} - s_\textbf{B})}\,,
\end{equation}
satisfying 
\begin{equation}
   T_\textbf{B} < \overline{T}_\textbf{BC} \leq T_{H}\,.
   \label{conditionTBC}
\end{equation}
Defining the heat via the entropic integral guarantees Eq.~\eqref{conditionTBC}, so the process is manifestly consistent with the second law.

The difference between $\Delta e_\textbf{BC}$ and $q_\textbf{BC}$ represents the energy stored in the pair correlations, $w_\textbf{BC}^{\rm int}$. In other words, since our engine operates via phase transitions during which the correlation energy varies, part of the energy is consumed as internal work. This is related to a change in the system's free energy per particle, $w^{\mathrm{int}} = \Delta F/N$, that is, the maximum energy available for non-mechanical work. This change in free energy reflects the energy exchange between the gas's subsystems, correlated pairs and unpaired fermions, that is required to keep the interaction parameter fixed at constant density. The internal work thus arises from the adjustment of the chemical potential due to the reorganization of the phases in response to the increase in temperature.

\subsubsection*{\texorpdfstring{\textbf{3. Statistical hot isotherm} \(\textbf{\textup{C}}\to \textbf{\textup{D}}\)}{\textbf{3. Statistical hot isotherm C->D}}}

At fixed \(n_2\) and while the system remains in contact with the hot reservoir at temperature \(T_H\), the scattering length is tuned isothermally, so that \(c_F\) changes from \(c_F^{\ell}\) to \(c_F^h\), where $c_{F,\textbf{D}}=c_{F}^h$. The final state is
\begin{equation}
    \textbf{D}=\left(\theta_H^{}~,~c_F^{h}\right)_{n_2}\,.
    \label{eq:state_D_cycle}
\end{equation}
This stroke corresponds to the fermionization of the gas driven by a statistical transformation of the system. The total internal energy change is
\begin{equation}
\begin{aligned}
    u_\textbf{CD}
    &=
    e_\textbf{D}(T_H,c_F^h) - e_\textbf{C}(T_H,c_F^\ell)
    \,,
\end{aligned}
\label{eq:DeltaeCD}
\end{equation}
however this comprises of two distinct contributions. The first is the heat exchanged with the hot reservoir caused by the change in the interaction parameter
\begin{equation}
\begin{aligned}
    q_\textbf{CD}^{s}
    =T_H^{}
    \left[
        s^\star_\textbf{D}(T_H^{},c_F^{h})
        -
        s^\star_\textbf{C}(T_H^{},c_F^{\ell})
    \right]\,.
\end{aligned}
\label{eq:qCDs}
\end{equation}
\\
\noindent
This heat flow is the result of work being done during the statistical change, which can be quantified through the difference between $u_\textbf{CD}$ and $q_\textbf{CD}^{s}$
\begin{equation}
    w_\textbf{CD}^{s}
    =
    u_\textbf{CD}
    -
    q_\textbf{CD}^{s}\,.
    \label{eq:wCDs}
\end{equation}
We refer to these two terms as statistical heat and statistical work, respectively. It should be noted that the density during this process is constant; therefore, the statistical stroke described here is a strictly isothermal and isochoric process.

\subsubsection*{\texorpdfstring{\textbf{4. Adiabatic expansion} \(\textbf{\textup{D}}\to \textbf{\textup{E}}\)}{\textbf{4. Adiabatic expansion D->E}}}

The gas is isolated from the baths and expanded from \(n_2\) to \(n_1\) at fixed \(c_F^{h}=c_{F,\textbf{E}}\). The final state is
\begin{equation}
    \textbf{E}=\left(\theta_\textbf{E}~,~c_F^{h}\right)_{n_1}\,.
    \label{eq:state_E_cycle}
\end{equation}
Just like the complementary stage, the isentropic expansion implies that $\theta_\textbf{D}=\theta_\textbf{E}$ and $T_\textbf{E}$ is fixed by
\begin{equation}
    s^\star_\textbf{E}(T_\textbf{E},c_F^{h})
    =
    s^\star_\textbf{D}(T_H,c_F^{h})\,.
    \label{eq:isentropic_DE}
\end{equation}
Since \(q_\textbf{DE}=0\), the work performed is
\begin{equation}
\begin{aligned}
    w_\textbf{DE}
    &=
    e_\textbf{E}(T_\textbf{E},c_F^h) - e_\textbf{D}(T_H,c_F^h)\,.
\end{aligned}
\label{eq:wDE}
\end{equation}

\subsubsection*{\texorpdfstring{\textbf{5. Isochoric cooling} \(\textbf{\textup{E}}\to \textbf{\textup{F}}\)}{\textbf{5. Isochoric cooling E->F}}}

The gas couples until equilibrium with the cold reservoir at temperature $T_{L}$, with \(n_1\) and \(c_F^{h}=c_{F,\textbf{F}}\) kept constant. The final state is
\begin{equation}
    \textbf{F}=\left(\theta_L^{}~,~c_F^{h}\right)_{n_1}\,.
    \label{eq:state_F_cycle}
\end{equation}
As in the heating stroke, the energy exchanged with the bath contains a contribution from the reconfiguration of internal correlations
\begin{equation}
    \Delta e_{\textbf{EF}}=q_{\textbf{EF}}+w_{\textbf{EF}}^{\rm int}\,.
\end{equation}
The heat extracted from the system is therefore computed via the entropic integral
\begin{equation}
q_{\textbf{EF}} =\frac{1}{n_1} \int_{s_\textbf{E}}^{s_\textbf{F}} T\,ds=\frac{1}{n_1} \int_{T_\textbf{E}}^{T_\textbf{F}} T\,\left( \frac{\partial s}{\partial T} \right)_{n_1, c_F^h}dT.
 \label{eq:qef}
\end{equation}
The mean entropic temperature of this stroke is
\begin{equation}
    \overline{T}_{\textbf{EF}} =\frac{q_{\textbf{EF}}} {(s_\textbf{F} - s_\textbf{E})}\,,
\end{equation}
satisfying 
\begin{equation}
   T_{L} \leq \overline{T}_\textbf{EF} < T_\textbf{E}\,.
\end{equation}

\subsubsection*{\texorpdfstring{\textbf{6. Statistical cold isotherm} \(\textbf{\textup{F}}\to \textbf{\textup{A}}\)}{\textbf{6. Statistical cold isotherm F->A}}}

Finally, at fixed \(n_1\) and fixed physical temperature \(T_L\), the scattering length is tuned isothermally from \(c_F^h\) back to \(c_F^{\ell}\), closing the cycle. This is the bosonization stroke. The heat exchanged with the cold reservoir is
\begin{equation}
\begin{aligned}
    q_\textbf{FA}^{s} = T_L
    \left[
        s^\star_\textbf{A}(T_L^{},c_F^{\ell})
        -
        s^\star_\textbf{F}(T_L^{},c_F^{h})
    \right]~~,
\end{aligned}
\label{eq:qFAs}
\end{equation}
while
\begin{equation}
\begin{aligned}
    u_\textbf{FA} = e_\textbf{A}(T_L,c_F^\ell) - e_\textbf{F}(T_L,c_F^h)~~,
\end{aligned}
\label{eq:DeltaeEF}
\end{equation}
and the statistical work performed is defined as
\begin{equation}
    w_\textbf{FA}^{s}
    =
    \Delta e_\textbf{FA}
    -
    q_\textbf{FA}^{s}~~.
    \label{eq:wFAs}
\end{equation}
Again, this is an isothermal-isochoric process.

\subsection{Engine Characterizations}
\label{subsec:cycle_balance_efficiency}

\begin{figure}[!t]
\centering
\includegraphics[width=\columnwidth]{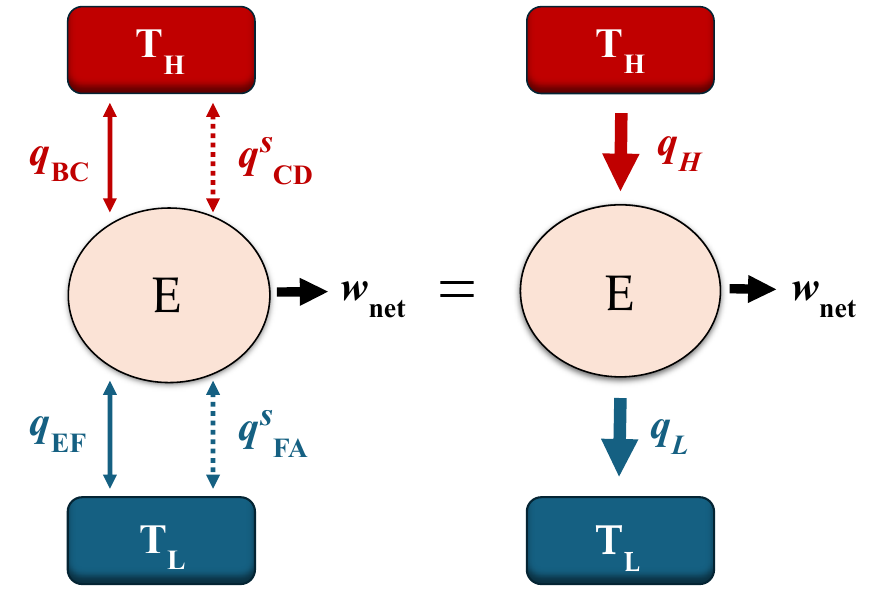}
\caption{Equivalent flow diagram of the statistical thermal cycle as a heat engine. The effective heats are resolved at the hot and cold reservoirs according to $q_{H} = q_{\textbf{BC}} + q_{\textbf{CD}}$ and $q_{L} = q_{\textbf{EF}} + q_{\textbf{FA}}$, respectively.}
\label{fig:Diagrama_de_flujo}
\end{figure}

One of the operational regimes of the STC is that of a heat engine, which we denote as the Statistical Thermal Engine (\textbf{ST-Engine}) and illustrate in Fig.~\ref{fig:Diagrama_de_flujo}. The ST-Engine operates by converting heat flow from the hot bath into usable work, with the net work produced by the engine defined as
\begin{equation}
   w_{\rm net} = q_{H}+q_{L}\,,
\label{eq:net_workfirst_law_cycle_2}
\end{equation} 
where $q_{H}$ and $q_{L}$ are the total heat from the hot and cold reservoirs, respectively
\begin{equation}
   q_{H}= q_\textbf{BC}+q_\textbf{CD}^s\,,~~q_{L}= q_\textbf{EF}+q_\textbf{FA}^s\,.
    \label{eq:first_law_cycle_2}
\end{equation}
The heat flow from each reservoir includes the contribution of both the isochoric processes ($q_\textbf{BC}$ and $q_\textbf{EF}$) along with the statistical heat ($q^s_\textbf{CD}$ and $q^s_\textbf{FA}$) created during the fermionization and bosonization processes. 

An important point to note is that the direction of the heat flow for both contributions can change, depending on how much the gas density is changed following the isentropic strokes, and the difference between the interaction parameters $c^{\ell}_F$ and $c^{h}_F$ during the isothermal strokes. Therefore, to characterize the operational regime of the heat engine we are only concerned with the total heat flow from both reservoirs, requiring that
\begin{equation}
    q_{H}>0~~,~~
    q_{L}<0~~,~~
    w_{\rm net}>0.
 \label{eq:engine_conditions_cycle}
\end{equation}
Other combinations of signs correspond to operational regimes that differ from those of a heat engine, which are discussed in Sec.~\ref{subsec:other_cycles}. The efficiency of the ST-Engine is similarly defined in terms of the total heat from the hot reservoir
\begin{equation}
    \eta_{\text{ST}}
    =
    \frac{w_{\rm net}}{q_{H}}
    =
    \frac{q_{H}+q_{L}}{q_{H}}
    =
    1+\frac{q_{L}}{q_{H}}\,.
    \label{eq:eta_cycle}
\end{equation} 

Finally, it is instructive to compare the efficiency of the ST-Engine to two well-known limits. The first is the Otto efficiency 
\begin{equation}
    \eta_{\text{Otto}}=1-r_c^{-2/3}\,,
\end{equation}
which describes Otto cycles and is thus expressed only in terms of the gas compression ratio, $ r_c = n_2/n_1$. The second is the Carnot efficiency
\begin{equation}
    \eta_C=1-r_t\,,
\end{equation}
which is the upper bound for all heat engines and depends solely on the ratio of the two reservoir temperatures, $ r_t = T_L/T_H$.

\section{Results and discussion}
\label{sec:results}

We focus on cycles in which the interaction parameter in the initial branch is set to $c_F^\ell=1.5$, and the target branch $c_F^h$ is scanned through the crossover for values between $-1.5$ and $1.5$. Examining this range allows us to see how the cycle's response changes when the working medium is driven from the molecular BEC side of the crossover towards the fermionic regime.

The cold temperature is fixed at $T_L=0.1E_{F,1}$, ensuring a condensed gas at low temperatures, as the critical temperature at $c_F=1.5$ within our FRG framework is $T_c^{\mathrm{BEC}}=0.26E_{F}$ (see appendix~\ref{app:renormalized_thermodynamics}). The thermal and compression ratios play distinct roles: \(r_t\) fixes the hot reservoir temperature, whereas \(r_c\) changes the Fermi energy scale between the two density branches.

\subsection{ST-Engine performance without compression and expansion stages}
\label{subsec:st_engine_no_compression}

\begin{figure*}[t]
\centering
\includegraphics[width=1\textwidth]{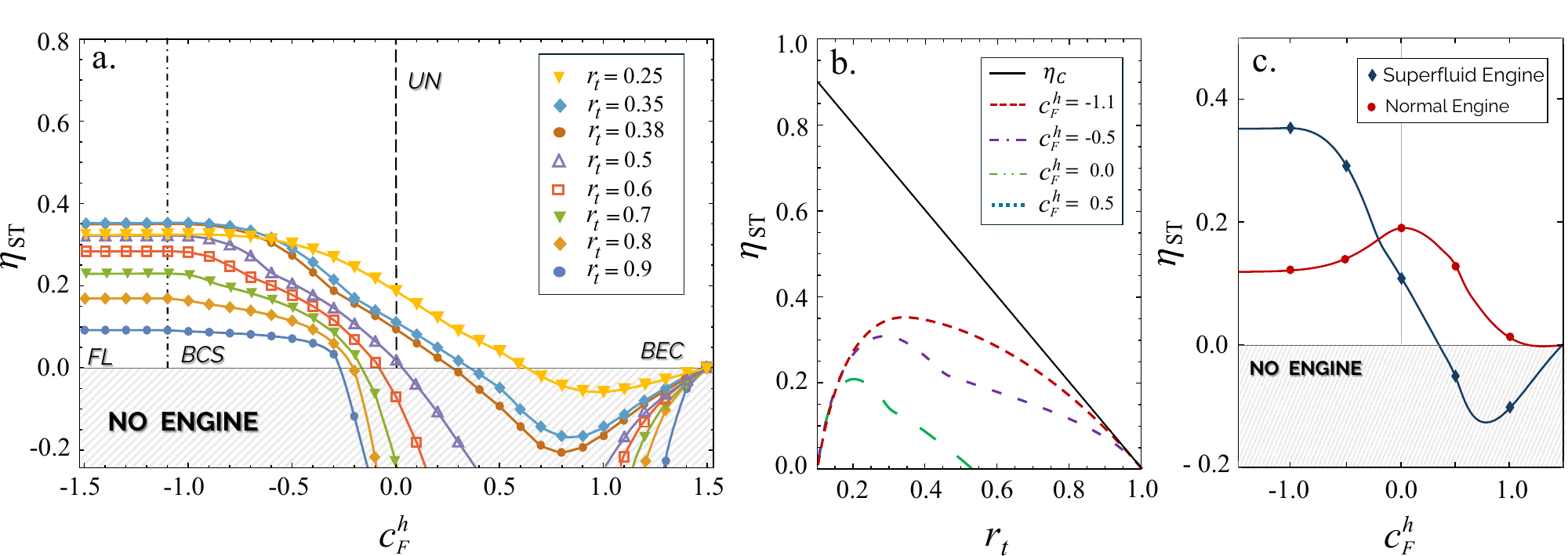}\caption{\label{fig:Efficiency_ST_Engine_1}~Efficiency at the BCS--BEC crossover, $\eta_{\text{ST}} = w_{net}/| q_H|$, for fixed compression ratio $r_c = 1$.~(a) The target branch $c_F^{h}$ of the cycle is moved from the BEC side, to unitarity (UN, dashed line), the BCS side and the Fermi liquid (FL) regime for different \(r_t=T_L/T_H\).~(b) Dependence of $\eta_{\text{ST}}$ on $r_t$, compared with the Carnot efficiency $\eta_C$. All calculations in (a) and (b) use $c_F^{\ell} = 1.5$ and $\theta_L = 0.1$. (c) Superfluid Engine $(\theta_L=0.1$ $r_t=0.35)$ is compared with Normal Engine $(\theta_L=0.3$, $r_t=0.59)$, with $T_c^{ \mathrm{BEC}}=0.26E_F$ for $c_F=1.5$.
}
\end{figure*}

Firstly, we analyze the cycle when statistical strokes are the only source of useful work. In this case, $r_c=1$, which implies that states $\textbf{A}$ and $\textbf{B}$ are the same, as well as $\textbf{D}$ and $\textbf{E}$, since 
\begin{equation}
    n_2=n_1\equiv n\,,~~ E_{F,2}=E_{F,1}\equiv E_F\,.
    \label{eq:equal_density_results}
\end{equation}
The adiabatic compression and expansion stages, therefore, become trivial, and the six-stroke cycle is reduced to an effective four-stroke cycle composed of two thermal isochoric and two isothermal statistical processes. This case is equivalent to a statistical Stirling cycle (SSC), but with the usual volume-changing isothermal transformations replaced by interaction-driven strokes performed at fixed temperature and fixed density.

Fig.~\ref{fig:Efficiency_ST_Engine_1}(a) displays the efficiency of the ST-Engine as a function of the target parameter $c_F^h$ for different temperature ratios $r_t$. For $0<c_F^h \leq 1.5$, the target branch remains on the BEC side, with $c_F^h=0$ being unitarity, while for negative values $c_F^h$ the target state crosses to the fermionic side. Within the latter, for $-1.1\leq c_F^h<0$ the \textbf{F} state lies within the superfluid BCS regime, while for $-1.5\leq c_F^h<-1.1$ the target branch is completely in the Fermi liquid (FL) regime. This is because at $c_F =-1.1$ the critical temperature on the BCS side ($T_c^{ \mathrm{BCS}}$) coincides with $T_L$. The shaded region identifies the parameter values for which the heat engine conditions of Eq.~\eqref{eq:engine_conditions_cycle} are not simultaneously satisfied; therefore, only the unshaded region is considered genuine ST-Engine operation. 

A pronounced asymmetry appears across the crossover. When both statistical strokes connect branches on the BEC side, the STC links thermodynamically similar paired states, and the work output remains small or becomes negative over a broad range of parameters (right side of the panel). In contrast, when $c_F^h$ is moved across unitarity and towards the fermionic side (left side of the panel), the statistical strokes connect states with more distinct entropy and internal energy structures (see appendix~\ref{app:renormalized_thermodynamics}). This increases the statistical contribution to the net work and broadens the engine operating window.

The efficiency converges to a constant value for $c_F^h \geq 1.1$, reaching its maximum when the state $\textbf{F}$ of the cold statistical branch is located at $T_\textbf{F} \simeq T_c^{\mathrm{BCS}}$, and the state $\textbf{C}$ of the hot statistical branch at $T_\textbf{C} \simeq 0.285\,E_F$ ($r_t = 0.35$). An increase in $T_H$ (to smaller $r_t$) improves the thermodynamic contrast, leading to greater heat absorption in $\textbf{B}\to \textbf{C}$. However, if $T_H \gg T_c^{\mathrm{BEC}}$, the pairing correlations decay and there is less statistical heat in $\textbf{C}\to \textbf{D}$, and thus the crossover advantage disappears. Therefore, the engine efficiency reflects a balance between thermal bias and the persistence of correlated pairs.

In Fig.~\ref{fig:Efficiency_ST_Engine_1}~(b), the efficiency behavior as a function of $r_t$ is shown more clearly. The black line represents the Carnot limit, and, as can be observed, the calculated efficiencies of the ST-Engine remain below this limit throughout the engine domain. This behavior confirms that the thermodynamic construction in the previous sections provides a consistent performance measure. The efficiency maxima of each curve, together with the observed deviation between $\eta_{\mathrm{ST}}$ and $\eta_{C}$ as the hot bath temperature increases, is, as mentioned, due to the decay of correlated pairs. Thus, the crossover does not provide an intrinsic thermodynamic advantage, but rather a controllable mechanism for converting changes in correlations into useful work. Since engine operation relies on the large difference in entropy between the different sides of the crossover, the performance deteriorates significantly when the hot branch reaches the Fermi temperature, $T_H \to T_{{F}}$ at $r_t=0.1$, and the gas becomes increasingly classical.

\medskip
\noindent\textit{Superfluid Engine $\&$ Normal Engine.}
\medskip

While the STC we have described exploits the superfluid phase to operate efficiently, it can still produce work above the critical temperature $T_c$. This is possible due to the pseudogap existing for $T_c < T < T^*$, with $T^*$ being the pseudogap temperature (see Ref.~\cite{Chen2005}). In this regime, preformed pairs retain the capacity to perform statistical work despite their lower coherence compared with the condensed gas. Therefore, we distinguish two operating regimes based on the phase the working medium traverses during the cycle.

A \textit{superfluid engine} is one in which at least one state of the cycle lies within the superfluid phase. It should be noted that this is not restricted to the initial state, since the trajectory may begin in the normal phase and subsequently enter the superfluid region. A \textit{normal engine}, by contrast, operates entirely within the normal phase and never crosses into the superfluid regime. Fig.~\ref{fig:Efficiency_ST_Engine_1}(c) compares the superfluid and normal engines. Something that characterizes and distinguishes both engines is that, for the superfluid engine, \(q_{\textbf{CD}}\) contributes more than \(q_{\textbf{BC}}\) to the total heat input. In the normal engine the opposite occurs, \(q_{\textbf{BC}}\) predominates over \(q_{\textbf{CD}}\). This implies that the efficiency is maximized at different temperatures, specifically at $r_t=0.35$ for the superfluid engine and $r_t=0.59$ for the normal one. This gives rise to different operating conditions.

In particular, we find that when operating at lower temperatures, the superfluid engine has increased efficiency when moving through the crossover towards the fermionic side (BCS side), but at the cost of reducing its operating range as a purely bosonic engine (BEC side). Conversely, we see that at higher temperatures the normal engine reaches its maximum efficiency at unitarity \(c_F^h = 0\) and that its operating window as an engine extends throughout the entire crossover. This behavior is explained by the change in the entropic structure of the system. The entropy peak shifts toward the BEC side at higher temperatures and, in turn, the entropic contrast between the bosonic and fermionic sides decreases (see appendix~\ref{app:renormalized_thermodynamics}, Fig.~\ref{fig:entropy_cF}), leading to a reduced maximum efficiency. 

\medskip
\noindent\textit{Statistical energy and maximum heat.} 
\medskip

Finally, we examine in greater detail the individual contribution of the statistical strokes, $\textbf{C}\rightarrow\textbf{D}$ and $\textbf{F}\rightarrow\textbf{A}$. This analysis provides physical information about the internal behavior of the ST-Engine. To illustrate the characteristic behavior of the statistical strokes, we consider a representative cycle with $ c_F^h = -0.4$. This configuration is chosen because it makes the relevant maxima more visible; the qualitative interpretation discussed below applies to the engine's operating region whenever the statistical stroke contributes more to the cycle than the isochoric stroke, i.e., for $T_L\ll T_c$.

\begin{figure}[!t]
\centering
\includegraphics[width=\columnwidth]{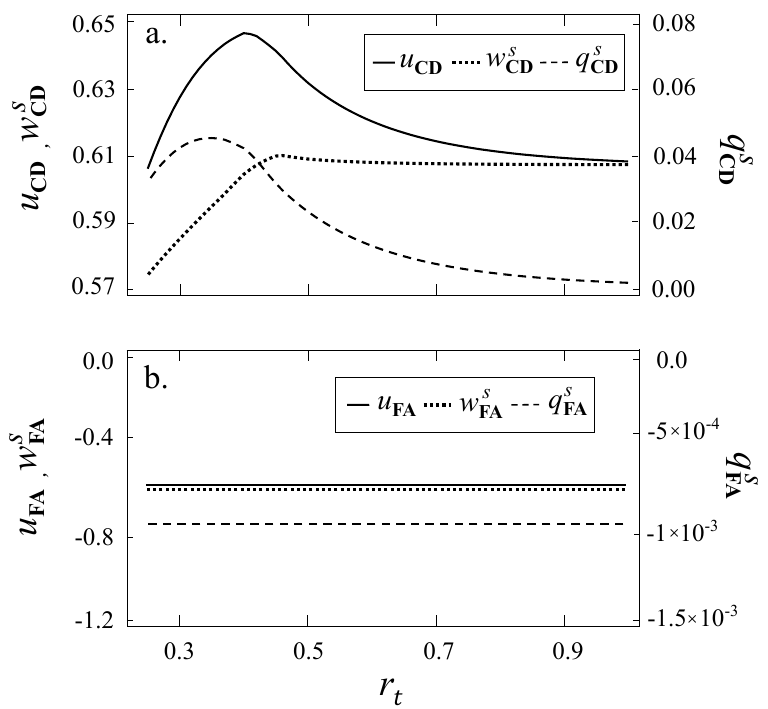}
\caption{\label{fig:Stadistical_Energy} First law balance. (a) The statistical hot stroke (\textbf{C} $\to$ \textbf{D}). (b) The statistical cold stroke (\textbf{F} $\to$ \textbf{A}). The internal energy \(u^{s}\), statistical heat \(q^{s}\) and statistical work \(w^{s}\) are shown as functions of  $r_t = T_L/T_H$. The displayed calculations use $r_c=1$, $c_F^{\ell} = 1.5$, $c_F^{h} = -0.4$ and $\theta_{L} = 0.1$. }
\end{figure}

Fig.~\ref{fig:Stadistical_Energy} shows the first-law balance of the hot ($\textbf{C}\to \textbf{D}$) and cold ($\textbf{F}\to \textbf{A}$) statistical strokes. The most relevant feature is observed in the hot branch, where the absorbed heat ($q_\textbf{CD}^s$) reaches its maximum at $r_t = 0.35$. This indicates that, in the regime where the statistical effects dominate the cycle's energy response, the maximum efficiency is strongly correlated with the maximum heat supplied by the statistical hot stroke.

A second important feature observed is that the maximum of $u$, $q^s$, and $w^s$ in the hot branch do not occur at the same high temperature. Instead, the characteristic reduce temperatures satisfy
\begin{equation}
\theta_{q_{\max}^{s}}^{} > \theta_{u_{\max}}^{} > \theta_{w_{\max}^{s}}^{}~~.
\end{equation}
Therefore, the condition that maximizes the heat absorbed during the statistical change is not the same as the condition that maximizes the statistical work, which manifests as a trade-off between efficiency and useful work. That is, the operating point favoring maximum efficiency is slightly displaced from the one favoring maximum useful work.


\subsection{ST-Engine performance with compression and expansion stages}
\label{subsec:st_engine_with_compression}

\begin{figure}[!h]
\centering
\includegraphics[width=\columnwidth]{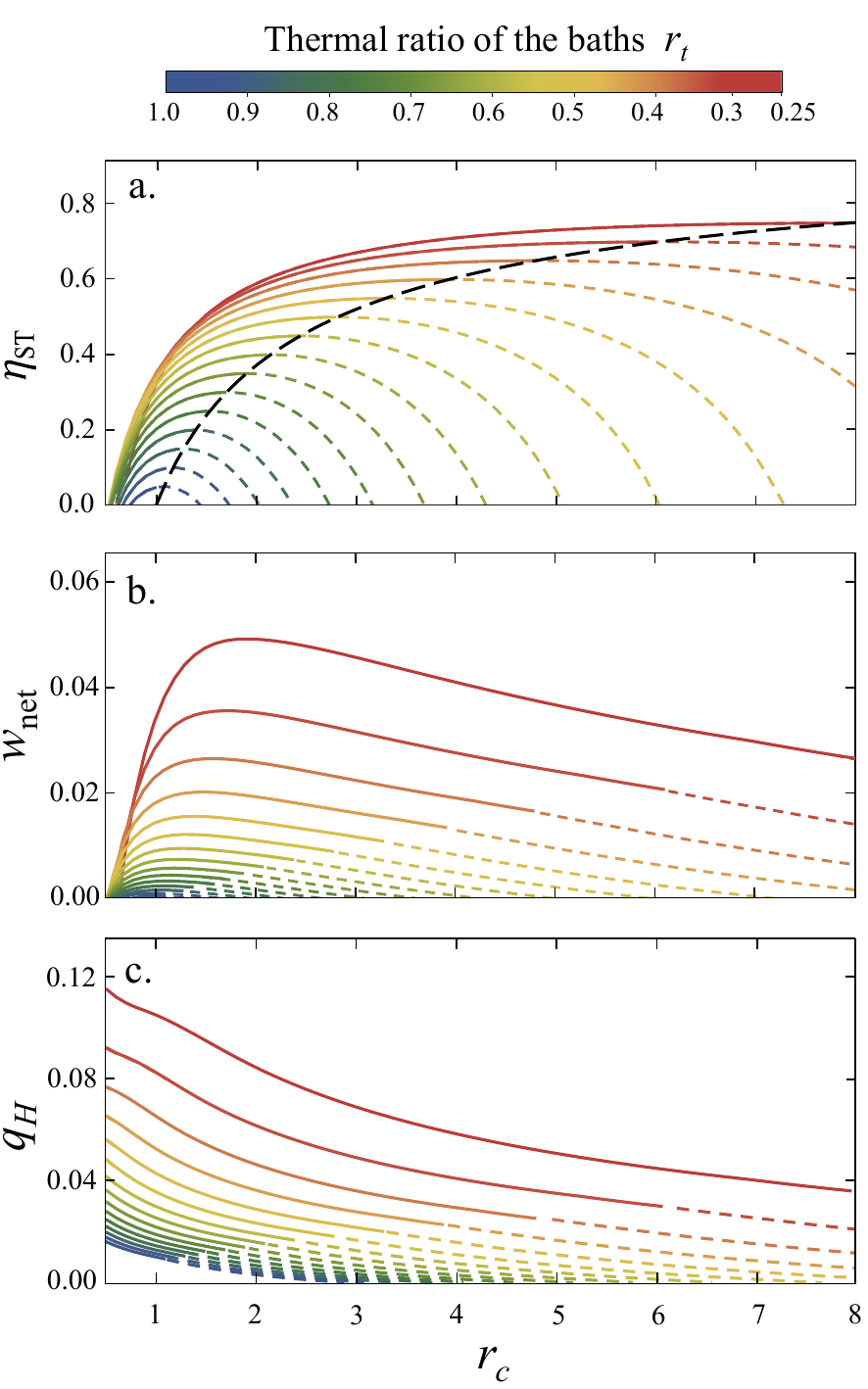}
\caption{(Color online). Performance of the ST-Engine as a function of the compression ratio $r_c$ for different values of $ r_t = T_L/T_H$, as indicated by the colorbar. (a) Efficiency $\eta_{\rm ST}=w_{\rm net}/q_{H}$. The black dashed line displays the Otto efficiency $\eta_{\rm Otto}=1-r_c^{-2/3}$. (b) Net work output $w_{\rm net}$. The work displays a maximum at an optimal compression ratio $r_c^{\rm opt}$. (c) Effective heat supplied $q_{H}$.  In all the panels, the solid colored curves consider cycles where $T_\textbf{B}<T_H$, while the dotted colored curves consider cycles where $T_\textbf{B}>T_H$. All the displayed calculations use $c_F^{\ell} = 1.5$, $c_F^{h} = -1.1$, and $\theta_{L} = 0.1$.}
\label{fig:Efficiency_ST_Engine_2}
\end{figure}

We proceed to examine the performance of the ST-Engine operating with the full STC, that is, including both compression and expansion stages. Fig.~\ref{fig:Efficiency_ST_Engine_2} summarizes the main engine quantities: efficiency ($\eta_{\text{ST}}$), net work ($w_{net}$), and total heat ($q_H$) supplied to the system from the hot bath, all as a function of $r_c$~. We set the target branch interaction parameter at $c_F^h=-1.1$, for which we attained the maximum efficiency at the crossover in Section \ref{subsec:st_engine_no_compression}. 

First, let us focus on the efficiency in Fig.~\ref{fig:Efficiency_ST_Engine_2}(a), which is shown for different temperature ratios $r_t$. In general, the maximum efficiency of the ST-Engine increases with the temperature of the hot bath $T_H$ (to smaller $r_t$), with the optimal compression ratio similarly increasing, as is usually found in Otto-like cycles. However, the ST-Engine is unique due to the presence of the statistical strokes, adding extra nuances to the engine performance. In particular, following the compression stroke $\textbf{A}\rightarrow\textbf{B}$, the temperature of the system at point $\textbf{B}$ can actually be larger than the hot bath, $T_\textbf{B}>T_H$, for sufficiently large $r_c$. We therefore denote these two regimes in the figure as solid curves when $T_\textbf{B}<T_H$, and dotted curves when $T_\textbf{B}>T_H$.

It may seem counter-intuitive that the ST-Engine operates when $T_\textbf{B}>T_H$, as in a standard Otto cycle, where the gas remains in the same phase throughout, the cycle would no longer operate as an engine, as no heat is absorbed by the hot bath due to $q_H<0$. However, in the ST-Engine, while $q_{\textbf{BC}}<0$, there can still be heat added to the system from the hot bath through the statistical heat $q_{\textbf{CD}}^s$ alone. In fact, as long as this is larger than the excess heat that is lost to the hot bath immediately following the compression stroke, we can ensure that $q_H>0$. The latter is shown in Fig.~\ref{fig:Efficiency_ST_Engine_2}(c), along with the net work output, as shown in Fig.~\ref{fig:Efficiency_ST_Engine_2}(b). 

The limit, $T_\textbf{B}=T_H$, is highlighted by the black dashed line in Fig.~\ref{fig:Efficiency_ST_Engine_2}(a) and denotes the boundary between these two thermal regimes. Here, $q_\textbf{BC}=0$ and $q_\textbf{EF}=0$, allowing us to realize a simplified cycle which has only four strokes: two isentropic ($\textbf{A}\rightarrow\textbf{B}$ and $\textbf{D}\rightarrow\textbf{E}$) and two isothermal ($\textbf{C}\rightarrow\textbf{D}$ and $\textbf{F}\rightarrow\textbf{A}$). This cycle represents the maximum efficiency attainable by the ST-Engine and is, in fact, where the Otto and Carnot efficiencies coincide, yielding $\eta_\text{max}=\eta_\text{Otto}=\eta_C$. The condition is achieved when $r_c^{-2/3}=r_t$, which is analogous to the relationship between the density and temperature ratios required for a conventional Carnot cycle. However, it should be noted that our engine operates within only two densities, $n_1$ and $n_2$, which are changed during the isentropic strokes. The isothermal strokes are isochoric, so the density is fixed, but instead we change the statistics through the interaction parameter. The resulting cycle combines features of both Otto and Carnot engines and may therefore be viewed as a hybrid Otto-Carnot cycle, which we term the Statistical Otto-Carnot Cycle (SOCC).

The SOCC is an intriguing case. In a conventional Otto cycle, the approach to the Carnot efficiency is typically accompanied by a vanishing cycle area and hence vanishing work output. In the present case, however, the statistical strokes permit a finite entropy change through variation of the interaction parameter $c_F$ across the statistical crossover. Consequently, the engine can continue to produce finite work even when operating at the Carnot point, as shown in Fig.~\ref{fig:Efficiency_ST_Engine_2}(b).

We note that the SOCC does not produce the maximum net work, and similar to the SSC presented in Sec.~\ref{subsec:st_engine_no_compression}, there is an efficiency-work trade-off. First, let us note that for $r_c<1$, the compression and expansion stages are swapped, and while the engine still operates, its performance rapidly decreases, eventually becoming inoperable around $r_c\approx0.5$. However, for $r_c>1$, the net work exhibits a maximum at an optimal compression ratio $r_{c}^{\mathrm{opt}}$. Initially, increasing $r_c$ enhances the difference between the energy scales of the compressed and expanded branches of the cycle, leading to a larger work output. However, for sufficiently large compression ratios, the population redistribution induced by the thermal reservoirs becomes less effective, causing the work output to decrease. Consequently, $r_{c}^{\mathrm{opt}}$ determines the location of the work maximum and is set by the cycle's compression, while the magnitude of this maximum is governed by $r_t$. This reflects the interplay between the thermal bias, which drives population redistribution, and the compression-induced shift in the Fermi energy, which eventually suppresses the bias.

\subsection{Operational regimes of the STC }
\label{subsec:other_cycles}

Up to this point, we have focused on the STC operations as a heat engine. The same cycle, however, can realize different thermodynamic modes because the signs of the heat exchanged with the two reservoirs can change as the interaction parameters are varied across the BCS–BEC crossover. 

\begin{figure}[!b]
\centering
\includegraphics[width=0.5\textwidth]{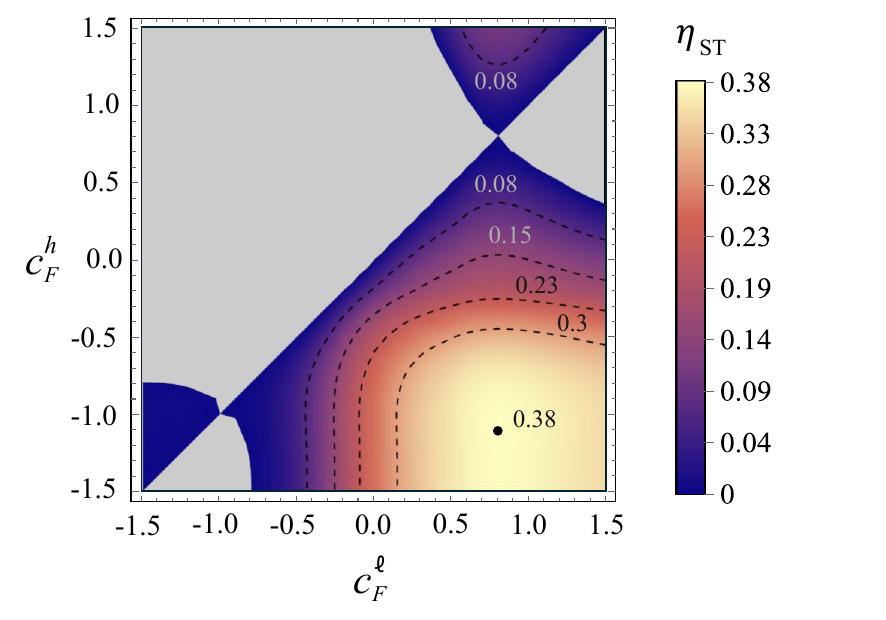}
\caption{\label{fig:operational_regimes} Efficiency $\eta_{\text{ST}}=w_{\rm net}/|q_H|$ of the statistical Stirling cycle (SSC) at $r_t=0.35$ with $\theta_L=0.1$ as a function of $c_F^\ell$ and $c_F^h$. The grey area represents where the cycle does not operate as an engine.}
\end{figure}

As a first examination, we consider the SSC ($r_c=1$) at fixed $r_t = 0.35$, and we map the efficiency across all possible values of $(c_F^\ell, c_F^h)$ in Fig.~\ref{fig:operational_regimes}. The colored regions indicate where the engine operates, with the largest efficiencies occurring when $c_F^{\ell}>c_F^{h}$, that is when traversing the crossover from the BEC to the BCS side during the statistical hot isotherm. Notably, its highest efficiency is attained at $(c_F^\ell=0.8,\, c_F^h=-1.1~;\ \eta_{\text{ST}}=0.38)$, which arises from the joint contribution of both branches. On the target branch, the efficiency peaks at $c_F^h=-1.1$ when $T_L \simeq T_c^{\mathrm{BCS}}$, as previously discussed. On the initial branch, for fixed $r_t=0.35$, the value $c_F^\ell=0.8$ corresponds to $T_H \simeq T_{\mathrm{BEC}}$. The simultaneous fulfillment of both conditions optimizes the statistical work and the heat supplied to the cycle, thereby enhancing the overall efficiency. 

On the opposite side of the parameter space for $c_F^{\ell}<c_F^{h}$, the engine mostly doesn't operate, which we indicate by the grey region. However, here different thermodynamic cycles can be realized according to the signs of $q_H$, $q_L$, and $w_{\rm net}$, as summarized in Table~\ref{tab:regimes}. In addition to the heat-engine regime, the STC can operate as a refrigerator, an accelerator, or a heater. In the accelerator regime, external work is supplied to enhance the natural transfer of heat from the hot to the cold reservoir, whereas in the refrigerator regime, external work drives heat from the cold reservoir toward the hot one. In the heater regime, the supplied work is ultimately released as heat to both reservoirs.
\begin{figure*}[t]
\centering
\includegraphics[width=1\textwidth]{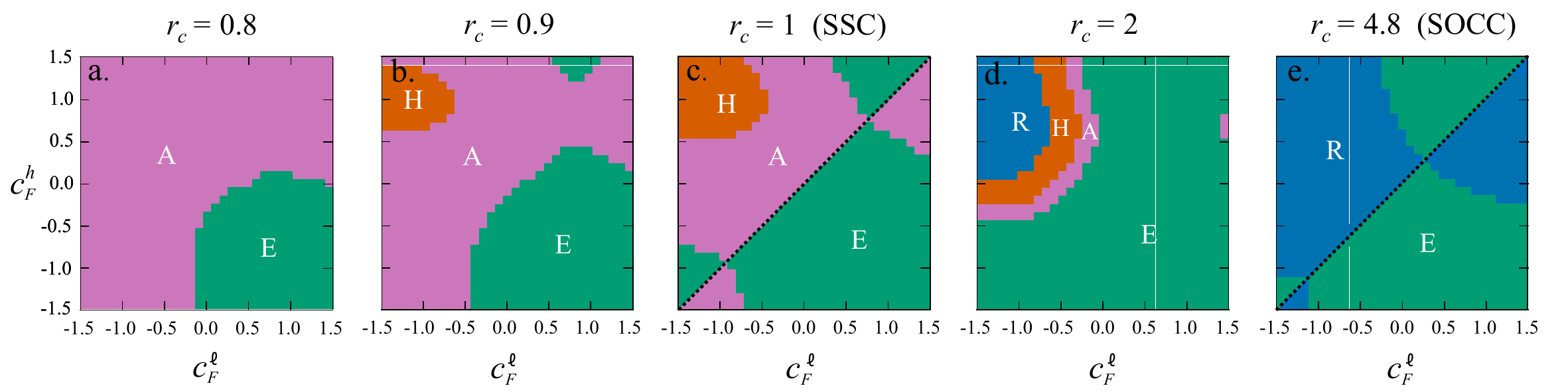}
\caption{\label{fig:operational_regimes_2}Operational regimes of the STC at the BCS--BEC crossover, with $r_t=0.35$ and $\theta_L=0.1$. The diagram highlights the domains where the cycle functions as an engine (E~-~green), refrigerator (R~-~blue), accelerator (A~-~pink) or heater (H~-~red), showing the machine's continuous tunability via the compression ratio. The black dotted line indicates where $w_\text{net}=0$ for the SSC and SOCC.}
\end{figure*}

The domains of these different processes are shown in Fig.~\ref{fig:operational_regimes_2} for different values of the compression ratio. Their behavior can be understood from the temperatures generated by the two isentropic density-changing strokes. Since the reduced temperature remains constant along these strokes, but the effective temperature changes with the compression ratio, we have
\begin{equation}
    T_\textbf{B}=T_L r_c^{2/3},
    \qquad
    T_\textbf{E}=T_H r_c^{-2/3}.
    \label{eq:isentropic_temperatures_operational}
\end{equation}
For $r_c<1$ (Fig.~\ref{fig:operational_regimes_2}(a) and Fig.~\ref{fig:operational_regimes_2}(b)), the density ordering is reversed: $\textbf{A}\rightarrow \textbf{B}$ is an isentropic expansion, whereas $\textbf{D}\rightarrow \textbf{E}$ is an isentropic compression. Consequently, $T_\textbf{B}<T_L$ and $T_\textbf{E}>T_H$, and the subsequent thermal strokes must bridge larger temperature intervals. This reversed orientation strongly suppresses the extraction of useful work. Accordingly, the accelerator regime occupies most of the interaction plane for strongly inverted density ratios, while heat-engine regions emerge progressively as $r_c$ approaches unity. Limited heater regions also appear because the statistical heat exchanged during the interaction-driven strokes can reverse sign across the crossover.

At $r_c=1$ (Fig.~\ref{fig:operational_regimes_2}(c)), the density-changing strokes become trivial, $\textbf{A}=\textbf{B}$ and $\textbf{D}=\textbf{E}$, and the STC reduces to the SSC. There is therefore no compression--expansion work channel, and the interaction-driven statistical transformations provide the externally controlled work mechanism. The coexistence of engine, accelerator, and heater sectors in this limit results entirely from the interaction dependence of the entropy and internal-energy differences between the two statistical branches.

For $r_c>1$(Fig.~\ref{fig:operational_regimes_2}(d) and Fig.~\ref{fig:operational_regimes_2}(e)), the usual compression--expansion ordering is recovered. Increasing $r_c$ raises $T_\textbf{B}$ and lowers $T_\textbf{E}$, progressively reducing the temperature intervals of both isochoric strokes. The relative importance of the statistical transformations therefore increases, and the operational pattern of the diagram changes continuously, favoring the engine and its reverse cycle as a refrigerator. In particular, for $r_{c}\simeq 4.8$, the condition $T_\textbf{B} \simeq T_\textbf{C}$ is satisfied, leading to the SOCC, in which the machine operates as a pure statistical engine or refrigerator.

Overall, Fig.~\ref{fig:operational_regimes_2} shows that the same working medium can be continuously adjusted between work extraction (E), work-assisted heat transport (A), and heating (H) simply by controlling the interaction endpoints. The refrigerator regime, however, cannot be reached through interaction tuning alone and requires a density-dependent modification of the Fermi energy.
\begin{table}[t!]
\caption{\label{tab:regimes} Operational regimes of quantum thermal machines based on the reservoir-resolved heats \(q_{H}\) and \(q_{L}\), and the total net work \(w_{\rm net}\). Positive \(w_{\rm net}\) denotes useful work delivered by the working medium, while negative \(w_{\rm net}\) denotes work supplied to the working medium.}
\begin{ruledtabular}
\begin{tabular}{lccc}
Device & \(q_{H}\) & \(q_{L}\) & \(w_{\rm net}\) \\
\hline\\[-1.6mm]
Engine (E)        & \(\geq 0\) & \(\leq 0\) & \(\geq 0\) \\
Refrigerator (R)  & \(\leq 0\) & \(\geq 0\) & \(\leq 0\) \\
Accelerator (A)   & \(\geq 0\) & \(\leq 0\) & \(\leq 0\) \\
Heater (H)        & \(\leq 0\) & \(\leq 0\) & \(\leq 0\) \\
\end{tabular}
\end{ruledtabular}
\end{table}

\section{Conclusions}
\label{sec:conclusions}

In this work, we proposed a statistical thermal cycle (STC) based on a strongly interacting Fermi gas across the BCS-BEC crossover. Using a non-perturbative FRG framework, we characterized the thermodynamics of the system throughout the crossover and across the superfluid-normal transition. Our results demonstrate that interaction-driven changes in the statistical nature of the gas enable control of heat flows, efficient work extraction, and access to multiple operational regimes within a single working medium.

A central aspect of this work is that interaction-driven transformations across the BCS-BEC crossover constitute a thermodynamic resource in their own right. We show that statistical work can be extracted solely from changes in the statistical character of the gas and that this mechanism is strongest near the superfluid transition, where pairing correlations and thermal driving act cooperatively. When combined with compression strokes, the resulting statistical Otto-Carnot cycle reaches both Otto and Carnot efficiency limits while retaining finite work output, in contrast to conventional heat engines. In turn, conditions that maximize efficiency often differ from those that maximize work output, revealing a trade-off between these two performance metrics. 

More broadly, the results show that the nontrivial entropy structure of the crossover enables the same device to operate as an engine, refrigerator, heat accelerator, or heater. This highlights the potential of strongly interacting quantum gases as versatile working media for quantum thermodynamic applications, opening new possibilities for the design of quantum thermal machines based on strongly correlated matter.

\section{Acknowledgments}
S.H.L. acknowledges financial support from ANID–Subdirección de Capital Humano, Doctorado Nacional, Grant No. 21220168 (2022), as well as from the “Programa de Iniciación a la Investigación Científica” (PIIC) of UTFSM under grant No. 015/2024. F.J.P. acknowledges support from FONDECYT (Chile) under Grant No. 1250173. P.V. and F. I. acknowledge support from CEDENNA under Grant CIA No. 250002. P.V. also acknowledges support from FONDECYT (Chile) under Grant No. 1240582. T.F. acknowledges support from JSPS KAKENHI Grant No. JP23K03290 and JST Grant No. JPMJPF2221. We also thank UTFSM, PUCV, and OIST for their institutional and financial support.

\appendix

\section{Critical Temperature and State Variables across the BCS--BEC crossover}
\label{app:renormalized_thermodynamics}

In support of the main text, we present the fundamental thermodynamic structure underlying the statistical strokes, namely the critical temperature, entropy, and internal energy across the BCS--BEC crossover, obtained via the FRG approach. For comparison, we also refer the reader to Refs.~\cite{Haussmann2007ThermoBCSBEC,bulgac2008quantum} for calculations with other approaches. In the following figures, the dimensionless interaction parameter $c_F = (k_F a)^{-1}$ dictates the regime: $c_F < 0$ corresponds to the fermionic BCS regime, $c_F > 0$ denotes the bosonic BEC regime, and $c_F = 0$ marks the strongly correlated unitary limit. All state variables are normalized by the Fermi energy $E_F$ and expressed in natural units ($k_B = \hbar = 1$) to ensure a direct comparison across the phase diagram.

\begin{figure}[!b]
\centering
\includegraphics[width=\linewidth]{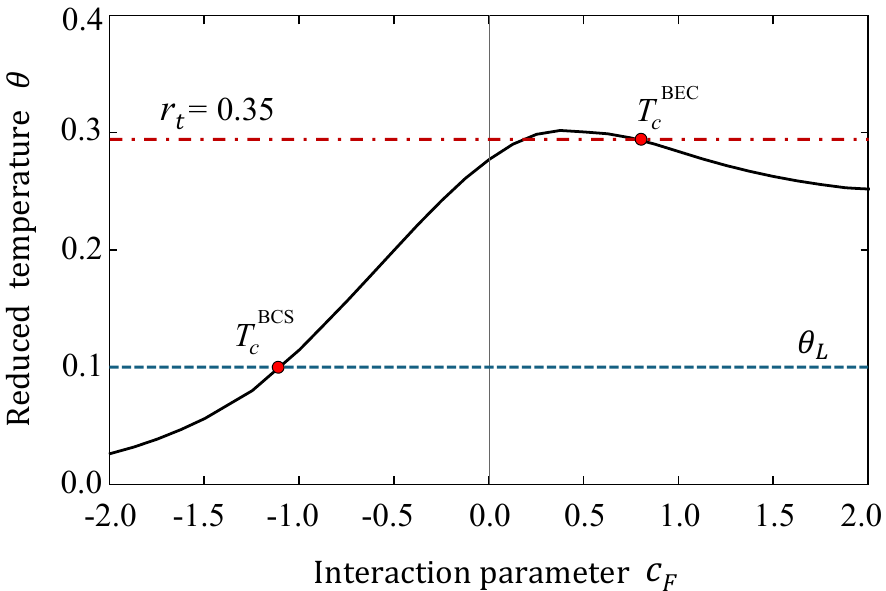}
\caption{Superfluid-to-normal phase transition temperature, with $\theta = T/E_F$ and $c_F = (k_F a)^{-1}$. 
Solid line: critical temperature $T_c/E_F$. 
Dash-dot line: level of maximum efficiency, $r_t = T_L/T_H$. 
Dashed line: cold reservoir temperature, $\theta_L=T_L/E_F$.}
\label{fig:phase_transition}
\end{figure}

Fig.~\ref{fig:phase_transition} depicts the superfluid-to-normal critical temperature $T_c$ mapped across the crossover within our FRG framework. The critical temperature rises from the exponentially suppressed BCS limit, crosses into the unitary limit, peaks on the near-BEC side, and then decreases. The initial temperature $\theta_L=0.1$ of the cycle is indicated by a dashed line, while the horizontal dash–dot line represents the reduced temperature ratio at which efficiency is maximized, $r_{t} \simeq 0.35$. The critical temperatures $T_{c}^{\mathrm{BEC}}$ at $c_F^\ell=0.8$ and $T_{c}^{\mathrm{BCS}}$ at $c_F^h=-1.1$ are also indicated, which correspond to the points where the statistical branches of states \textbf{C} and \textbf{F}, respectively, optimize the efficiency of the ST-Engine.

\begin{figure}[!t]
\centering
\includegraphics[width=\linewidth]{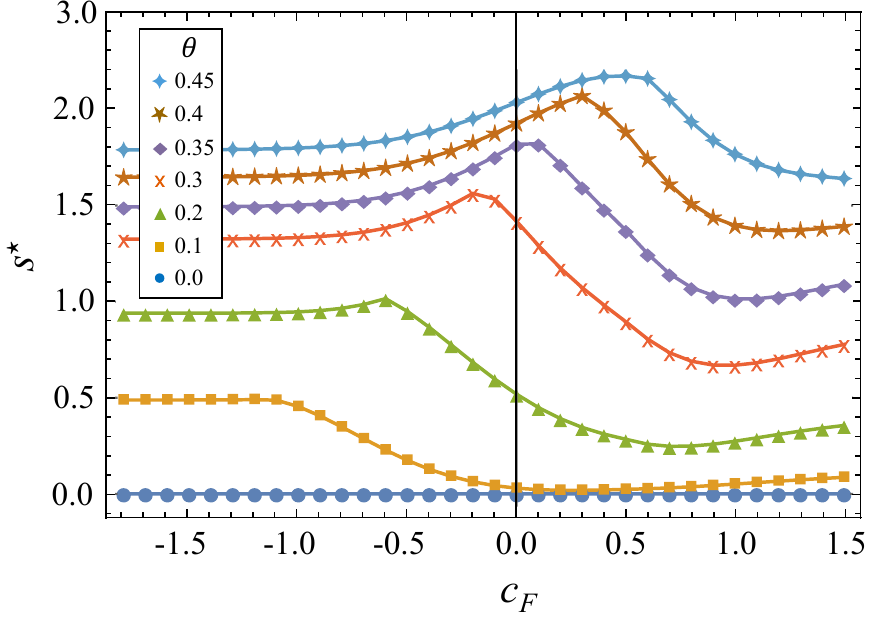}
\caption{Isothermal entropic lines, with $s^\star=S/N$ entropy per particle, $\theta=T/E_F$ and $c_F=(k_Fa)^{-1}$.}
\label{fig:entropy_cF}
\end{figure}

Fig.~\ref {fig:entropy_cF} shows the entropy per particle across the crossover under isothermal conditions. The distinctive asymmetry in the entropic structure between the fermionic and bosonic regimes determines both the magnitude and the direction of the statistical heat exchanged during the fermionization and bosonization of the gas, which vanishes at temperatures well above $T_c$. At low temperatures, the entropy decreases significantly on the BEC side as tightly bound bosonic molecules condense, thereby reducing the available thermal excitations. However, as the temperature approaches and exceeds $T_c$, the entropy exhibits a pronounced nonmonotonic behavior, shifting the entropy peak toward the BEC side due to the thermal dissociation of molecular dimers into their fermionic constituents.

\begin{figure}[!t]
\centering
\includegraphics[width=\linewidth]{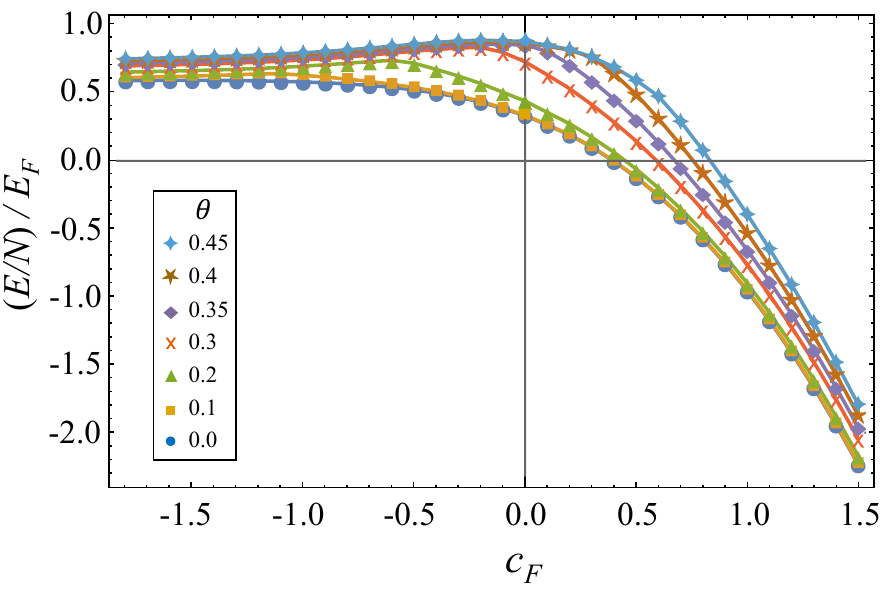}
\caption{Reduced energy per particle $(E/N)/E_F$ at different temperatures, with $\theta=T/E_F$ and $c_F=(k_Fa)^{-1}$}
\label{fig:energy_cF}
\end{figure}

Finally, Fig.~\ref{fig:energy_cF} illustrates the energy per particle. The steep monotonic drop towards the BEC side is dominated by the large negative binding energy $\epsilon_b = -1/(m a^2)$ of the condensed dimers. The substantial energy gradient between the weakly interacting BCS regime and the strongly bound BEC regime provides the internal energetic reservoir that the statistical strokes harness to output non-trivial generalized work.

\bibliography{bibfinal}

\end{document}